\documentclass[final,5p,times,twocolumn]{elsarticle}
\usepackage{amsmath}
\usepackage{amssymb}
\usepackage{graphicx}
\usepackage{dcolumn}
\usepackage{bm}
\usepackage{wasysym}
\usepackage{color}
\begin{document}

\title{Optimal confinement potential in quantum Hall droplets}

\author{E. T\"ol\"o and A. Harju}

\address{Helsinki Institute of Physics and Department of Applied Physics,\\
 Helsinki University of Technology, P.O. Box 4100, FI-02015 HUT, Finland}

\begin{abstract}
We find that the confinement potential of a few electron quantum dot
can be tuned to significantly increase the overlap with certain quantum 
Hall trial wave functions. Besides
manipulating inter-electron interaction, this approach may prove useful in 
quantum point contact experiments, which involve narrow constrictions.
\end{abstract}

\maketitle

\section{Introduction}

Fractional quantum Hall effect is observed in high-mobility
two-dimensional electron systems at low temperature and
strong perpendicular magnetic field as the interactions between
the electrons induce a gap in the bulk, and the transverse 
conductance is quantized universally as $G=\nu e^2/h$ 
with certain rational numbers $\nu$. Each filling fraction $\nu$
possesses quasiparticle excitations with characteristic fractional
charge and statistics. 

Recently, there has been a major technological interest in 
suspected non-abelian phases, in particular at filling fraction 
$\nu=5/2$, the quasiparticle properties of which could 
proposedly be applied in topologically protected quantum gates, 
robust against decoherence and local perturbations \cite{kitaev,nayak,bonderson}. The ongoing
experimental studies of the statistics of the $5/2$ state 
make use of quantum point contacts \cite{miller,dolev,radu,roddaro}, in which the electrons are 
confined away from the tunneling region. However, experiments suggest
that as the quantum point contact shrinks in size, the $5/2$ 
state eventually becomes unstable. Engineering the form of the confinement 
potential as to stabilize the non-abelian state in the quantum point contact 
region could therefore prove advantageous. 

In this paper, we find the confinement potentials that maximize the overlap of the
wave function of a few electron quantum Hall droplet with the Laughlin state at $\nu=1/3$ \cite{laughlin}
\begin{equation}
\Psi_{\mathrm{L}}(z_1,z_2,\ldots,z_N)=\textstyle{\prod}_{i<j} z_{ij}^3e^{-\sum_iz_i\bar{z}_i/2}\ ,
\end{equation}
where $z=x+iy$ and $z_{ij}=z_i-z_j$, and the Moore-Read state on the lowest Landau level (LLL) \cite{moore}
\begin{equation}
\label{pfaffian}
\Psi_{\mathrm{M-R}}^{\mathrm{LLL}}(z_1,z_2,\ldots,z_N)=\mathrm{Pf}\left(z_{ij}^{-1}\right)\textstyle{\prod}_{i<j} z_{ij}^2e^{-\sum_iz_i\bar{z}_i/2}
\end{equation}
and on the second lowest Landau level (SLL). The latter is obtained by applying the Landau level raising operator $\partial_z-\bar{z}/2$ to each electron coordinate in $\Psi_{\mathrm{M-R}}^{\mathrm{LLL}}$
and assuming the lowest Landau level of both spin types completely filled and inert.

The rest of the manuscript is organized as follows. The model and the optimization method is explained in Section 2. Section 3 contains the results and Section 4 the summary.

\section{Model and optimization}

The system is modelled by an effective-mass Hamiltonian 
\begin{equation}
\label{ham}
H=\sum^N_{i=1}\left[
 \frac{\left({\bf p}_i+\frac{e}{c} {\bf A}_i \right)^2}{2 m^*}
+ V(r_i)\right] +  \sum_{i<j}
\frac{e^2}{\epsilon r_{ij}}\ ,
\end{equation}
where $N$ is the number of electrons, and ${\bf A}$ is the vector potential
of the homogeneous  magnetic field ${\bf B}$ perpendicular to the sample plane.  
The material dependent parameters are $m^*=0.067m_e$, the effective mass 
of an electron, and $\epsilon=12.7$ (CGS), the dielectric constant of GaAs 
semiconductor medium. The confinement potential is expanded as
\begin{equation}
V(r)=\frac{m^*\omega_0^2r^2}{2}+\sum_{k=-n}^{k_{\mathrm{max}}}\frac{a_kn!m^*\omega^2l^2}{(k+n)!}(r/l)^{2k}e^{-(r/l)^2}\left(L_n^k((r/l)^2)\right)^2\ ,
\end{equation}
where the last term is motivated by the form of the single-particle basis. $k_{\mathrm{max}}$ is 
set to the maximum single-particle angular momentum (finite for a fixed total angular momentum), 
and $n$ is the Landau level index. $L_n^m$  are the generalized Laguerre polynomials, $L_0^m(x)=1$ and $L_1^m(x)=-x+m+1$. The optimal parameters $a_k$ are searched by a random walk whereby
new value is accepted whenever the resulting overlap is higher.

In the calculations, we set the confinement strength 
$\hbar\omega_0$ to $2\,{\rm meV}$ as its scaling should merely shift the 
ranges of magnetic fields for different phases. Lengths are written in units 
of oscillator length $l=\sqrt{\hbar/(m^*\omega)}$, where 
$\omega=\sqrt{\omega_0^2+(\omega_c/2)^2}$ and  $\omega_c=eB/(m^*c)$ is the 
cyclotron frequency. 
The lowest energy state of Eq. (\ref{ham}) with given angular momentum is solved 
by performing Lanczos diagonalization for the many-body Hamiltonian matrix constructed in the basis 
of a spin-polarized Fock-Darwin band corresponding to fixed $n$. This constitutes a Landau level projection, 
an approximation that is valid at the 
high magnetic field regime and for small enough deviation from the ideal
parabolic potential. The single-particle wave functions at Fock-Darwin band $n$ and angular momentum 
quantum number $m$ are written in oscillator units as
\begin{equation}
\label{fd}
\langle z|n,m\rangle=\sqrt{\frac{n!}{\pi (n+m)!}}z^mL_n^m(z\bar{z})e^{-z\bar{z}/2}\ , \ \ m\geqslant-n\ .
\end{equation}
The many-body Hamiltonian is written in second quantized formalism as
\begin{equation}
H=\sum_{i,j}(\delta_{ij}\epsilon_{i}^n+v_{ij}^n)c_i^{n\dagger}c_j^n+\sum_{i,j,k,l}v_{ijkl}^nc_i^{n\dagger}c_j^{n\dagger}c_l^nc_k^n\ ,
\end{equation}
where $\epsilon_{m}^n=(2n+1)\hbar\omega+l\hbar(\omega-\omega_c/2)$ are the single-particle energies in absence
of the optimization potential and $v_{mm'}^n=\langle n,m'|\sum_{k=-n}^{k_{\mathrm{max}}}\frac{a_kn!m^*\omega^2l^2}{(k+n)!}(r/l)^{2k}e^{-(r/l)^2}\left(L_n^k((r/l)^2)\right)^2|n,m\rangle$ are calculated analytically for 
the two lowest Landau levels (see Appendix). 
The non-trivial quantities are the interaction matrix elements $v_{ijkl}^n$, which are
calculated utilizing Tsiper's analytic formula \cite{tsiper_c}.

In the light of the numerical study \cite{wojs} and the 
recent tunneling experiment \cite{radu}, it seems plausible that the Landau level 
mixing, which breaks the particle-hole symmetry, favors the
particle-hole conjugated Moore-Read state termed the anti-Pfaffian \cite{levin}.
The potential could then be optimized in similar fashion for anti-Pfaffian
or any other candidate state.

\section{Results}

The obtained optimal overlaps are shown in Table \ref{ol}. For the Laughlin state, the initial overlaps 
are in general the highest, yet the lack of overlap decreases on average by 68 \%. %and the increase amounts to a factor of 1.01 to 1.02. 
In the lowest Landau level, the initial overlaps of the Moore-Read are quite modest but average 63 \% cut
in the lack of overlap makes a notable increase. On the second lowest Landau level the cut averages at about
84 \% and the overlaps are even higher.
%state increase more clearly by factor or 1.03 to 1.48 depending on the particle number and the Landau level.
Similarly increased overlaps have been reported earlier for modified Coulomb interactions whereby 
thickness or screening is added \cite{saarikoski_pf,peterson,tolo}. 

However, the relevance of few electron 
results for moderately large electron numbers remains uncertain. Firstly, the smaller the droplet, the larger
the finite size effects. Even if the overlap is quite low, the state may still belong to the same
topological phase as the trial wave function.
Secondly, despite the high overlap the new state could be energetically instable 
relative to states with different angular momentum than that of the trial wave function. Extrapolation of the
actual ground state to larger particle numbers then yields an instability to another filling fraction or a 
compressible state. Alternatively, one might therefore try to maximize the excitation gap of the ground
state phase keeping only a moderate overlap with the trial state.

\begin{table}[htb]
\begin{center}
\caption{The initial and optimized overlaps for the $1/3$ Laughlin state and the Moore-Read state at $\nu=1/2$ and $\nu=5/2$ with $N$ electrons.}
\label{ol}
\begin{tabular}{cccc}
\hline
$N$&$4$&$5$&$6$\\
\hline
$|\langle\Psi_{\mathrm L}|\Psi_i\rangle|$&0.979&0.985&0.982\\
$|\langle\Psi_{\mathrm L}|\Psi_f\rangle|$&0.995&0.995&0.993\\
\hline
$N$&$4$&$6$&$8$\\
\hline
$|\langle\Psi_{\mathrm{M-R}}^{\mathrm{LLL}}|\Psi_i\rangle|$&0.922&0.790&0.586\\
$|\langle\Psi_{\mathrm{M-R}}^{\mathrm{LLL}}|\Psi_f\rangle|$&0.963&0.933&0.866\\
\hline
$N$&$4$&$6$&$8$\\
\hline
$|\langle\Psi_{\mathrm{M-R}}^{\mathrm{SLL}}|\Psi_i\rangle|$&0.970&0.849&0.730\\
$|\langle\Psi_{\mathrm{M-R}}^{\mathrm{SLL}}|\Psi_f\rangle|$&0.997&0.978&0.935\\
\hline
\end{tabular}
\end{center}
\end{table}

\begin{figure*}[tb]
\includegraphics[width=2\columnwidth]{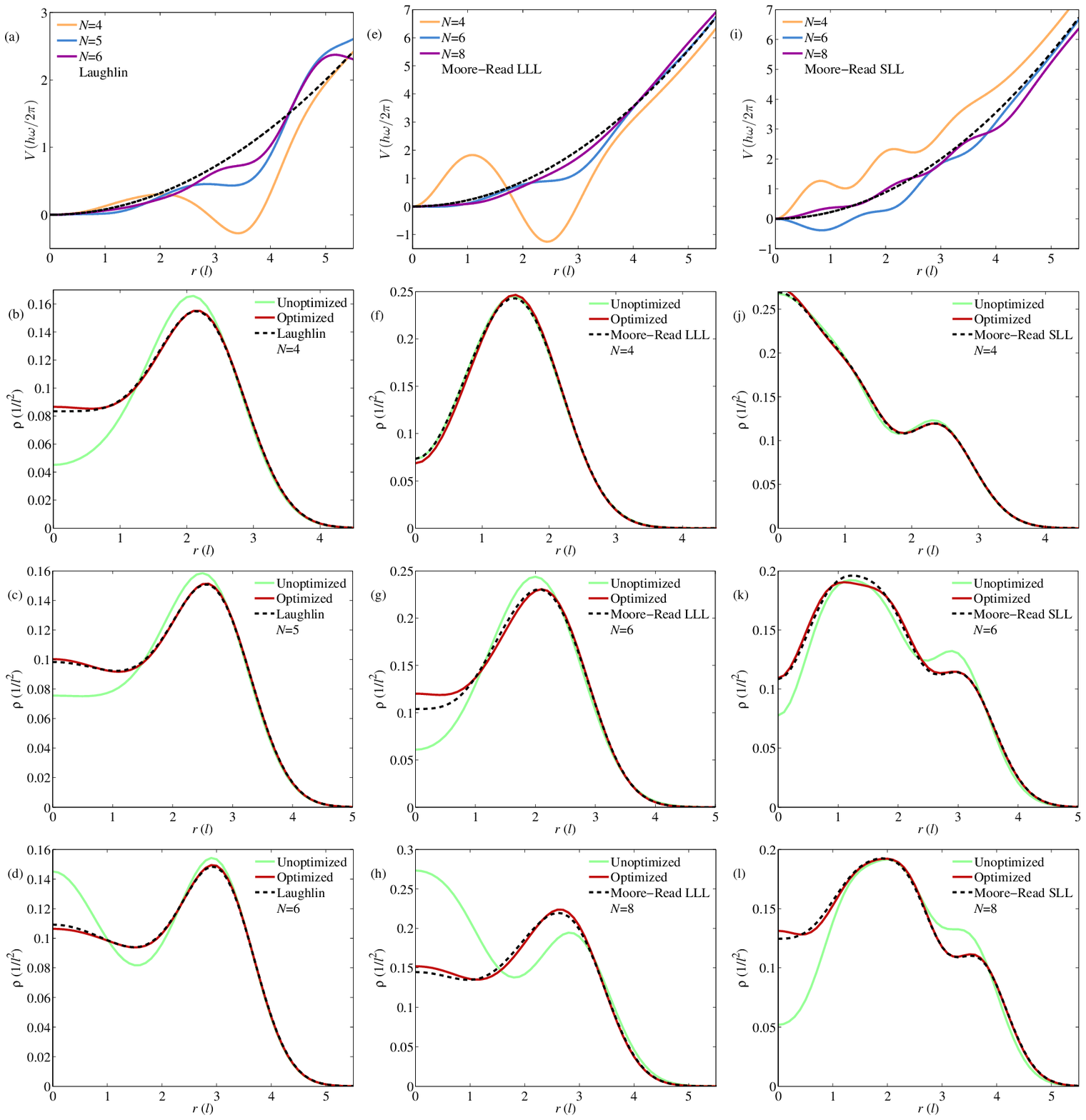}
\caption{The optimal confinement potential to maximize the overlap with 
(a) the Laughlin state with $\omega_0=\frac{2}{5}$,
(b) the Moore-Read state in the lowest Landau level, and
(c) the Moore-Read state in the second lowest Landau level with $\omega_0=\frac{2}{3}$,
where the dashed line corresponds to the parabolic potential. (e)-(m) the thus obtained elecronic densities compared to the trial
wave function and unoptimized potential.}
\label{kuva}
\end{figure*}

The modified potentials and their effect to the electron densities are presented in Figure \ref{kuva}.
The potentials are shifted by a constant as to ensure that $V(r)=0$ at $r=0$.
As seen in Figure \ref{kuva}(a),(e), and (i), the difference to the clean parabolic potential reduces as
the particle number is increased. This is likely to be related to smoothening of the electronic density
as the particle number is increased.
In the second Landau level there are more fluctuations in the
optimized potentials, since
the basis function there contain the first non-trivial generalized Laguerre polynomial factor leading
to an additional maximum in the single particle densities.
Looking at the electronic densities in Fig.~\ref{kuva}, the improvement in densities is
clearly visible in all cases except for Fig.~\ref{kuva}(f). Note, however, that due to the disc geometry, 
the optimal densities tend to be more optimal near the edge of the droplet where most of the density 
is accumulated. 

For the potentials optimized for the Laughlin state, in Fig.~\ref{kuva}(a), a common feature is the sharpness
of the potential compared to the parabolic potential near the edge of the droplet (see Fig.~\ref{kuva}(b-d)) around $r=4l$.
As noted in earlier works \cite{wan,tolo2}, this compression leads to a renormalization of the tunneling 
exponent \cite{chang,grayson} characterizing the non-Ohmic charge transport to the fractional quantum Hall edge.
In this case, the increase in the initially high overlaps seems marginal but the experimental consequences
may not be.

For the Moore-Read state in the lowest Landau level, the resulting densities (Fig.~\ref{kuva}(f-h)) match the
trial density almost as well as for the Laughlin state despite the notably lower initial overlaps.
The form of the potential near the edge is again slightly sharper than a parabolic.
A remarkable improvement occurs in the lowest Landau
level for 8 electrons (Fig.~\ref{kuva}(h)) where the overlap 
climbs up to 0.974. This raises question, whether filling fraction 
$\nu=1/2$ state with Pfaffian wave function (Eq. (\ref{pfaffian})) could be realized in a suitably tuned 
quantum Hall droplet. Our preliminary analysis,
however, indicates that the obtained state is instable against higher angular momentum states and
thus not immediately realized as the ground state. However, if slight variations of the potential landscape
are still allowed without destroying the state, not to mention a larger droplet with more electrons, 
this instability might be surmountable. A lowest Landau level non-abelian state would be especially welcome
since the Landau level mixing could then be well neglected,

In the second lowest Landau level, the Moore-Read state attains even larger overlaps and the densities
seem to match reasonably well (Fig.~\ref{kuva}(j)-(l)). However, the form of the optimal potentials is a 
bit more complicated with oscillatory behaviour.

\section{Summary}

We have calculated the confinement potentials that maximize the wave functional overlap of 
a few electron droplet with the $1/3$ Lauglin state and the Moore-Read state at 
$\nu=1/2$ and $5/2$. The overlaps were found to increase clearly, particularly for 
the larger electron numbers, and the obtained electron densities matched well with the fluctuations of
the trial densities. The Laughlin state and the Moore-Read state in lowest Landau level favored a sharpened
potential near the edge of the droplet.

\section*{Acknowledgements}

This study has been supported by the Academy of Finland through its Centres of Excellence Program (2006-2011). 
ET acknowledges financial support from the Vilho, Yrj\"o, and Kalle V\"ais\"al\"a Foundation of the Finnish
Academy of Science and Letters. We thank J. Suorsa for the interaction matrix elements formula.

\section*{Appendix}
In the lowest Landau level ($n=0$), the matrix elements $v_{mm}^n$  reduce to
\begin{equation}
v_{m'm}^0=\delta_{m'm}\textstyle{\sum}_k\frac{a_k}{2^{m+k+1}}\frac{(m+k)!}{m!k!}\ ,\quad m\geqslant0\ ,\  k\geqslant0\ .
\end{equation}
In the second lowest Landau level ($n=1$), we have
\begin{equation}
v_{m'm}^1=\delta_{m'm}\textstyle{\sum}_kh_{mk}a_k\ ,
\end{equation}
where we have $k\geqslant-1$ and $m\geqslant-1$, and the matrix $h_{mk}$ is given by
\begin{equation}
h_{mk}=\frac{(4 k^3 m - m^2 + 2 k m^2 + 6 k^2 m^2 - 2 m^3 - 4 k m^3 + m^4)(m+k)!}{2^{m+k+5}m!k!}
\end{equation}
for $m+k\geqslant0$, $h_{mk}=1/8$ for $m+k=-1$, and $h_{mk}=1/4$ for $m+k=-2$.

\end{document}